%% file: main.tex
\documentclass[reprint,aps,pra,twocolumn,floatfix,secnumarabic,hidelinks]{revtex4-2}

\input{style}

\input{commands}

\begin{document}
\title{\Title{}}

\input{title-config}

\begin{abstract}\noindent
    We generate multipartite entangled states of two, three and four matter qubits, where the entanglement is distributed over macroscopic distances via a photonic network link.
    Trapped-ion \Sr{88} qubits are entangled directly via the optical fibre link, and the entanglement is subsequently extended to \Ca{43} memory qubits co-trapped in each network node, using local mixed-species logic gates.
    We create remotely entangled \Sr{}-\Ca{} and \Ca{}-\Ca{} states, as well as mixed-species \ac*{GHZ} states of up to four qubits.
    We demonstrate storage of the remotely-entangled memory qubits for $\sim\SI{10}{\second}$, more than $100\times$ the creation time.
\end{abstract}

\maketitle
\noindent

\section*{Introduction}
Quantum networks enable the distribution of entanglement between physically separated systems.
This capability is critical for a variety of quantum technologies such as entanglement-based cryptography~\cite{ekert_quantum_1991}, distributed quantum computing~\cite{cirac_distributed_1999, jiang_distributed_2007, main_distributed_2025}, quantum sensing~\cite{guo_distributed_2020}, and metrology~\cite{komar_quantum_2014}.
Many of these applications are based on the ability to repeatedly generate multipartite entangled states between separated modules during the execution of a protocol.
Realising these applications requires network modules that can not only establish remote entanglement but also process and store quantum information locally between subsequent rounds of entanglement generation.
Therefore, integrating robust quantum memories with photonic interfaces into quantum networking modules is essential~\cite{kimble2008quantum,reiserer2015cavity, wehner2018quantum, pompili2021realization,hucul2015modular}.

The generation of multipartite entanglement across network nodes is a key milestone towards scalable architectures.
Multipartite entangled states underpin many advanced quantum networking protocols, such as networked quantum error correction, quantum secret sharing, and distributed sensing with entanglement-enhanced sensitivity~\cite{gottesman1999quantum,hillery1999quantum,komar2014quantum}.
Demonstrating such states in real networks is therefore essential to benchmark performance beyond bipartite entanglement.

At the same time, supporting multiple types of qubits within a quantum network is critical to building heterogeneous modular systems that combine the strengths of different types of qubits, such as optimal photonic coupling, high-fidelity logic, or long-lived memory~\cite{schmidt2005spectroscopy, tan2015multi, inlek2017multispecies, bruzewicz2019dual, hughes_benchmarking_2020}.
Cross-qubit entanglement opens the door to hybrid quantum networks that can integrate diverse platforms or future-proof evolving hardware.

Trapped ions are among the leading candidates for quantum information processing due to their long coherence times~\cite{wang_single_2021}, high-fidelity quantum gates~\cite{ballance_high-fidelity_2016, gaebler_high-fidelity_2016, hughes_benchmarking_2020, srinivas_high-fidelity_2021, clark_high-fidelity_2021, weber_robust_2024, loschnauer_scalable_2024, smith_single_2025}, and robust state preparation and measurement protocols~\cite{harty_high-fidelity_2014, an_high_2022, sotirova_high-fidelity_2024}.
Furthermore, trapped ions naturally interface with optical photons, and photonic interconnects between ion-based modules have demonstrated state-of-the-art networking capabilities, enabling the generation of remote entanglement between qubits separated by macroscopic distances~\cite{moehring2007entanglement, stephenson_high-rate_2020, krutyanskiy2023entanglement, saha_high-fidelity_2024}.

In this Letter, we present an elementary quantum network consisting of two mixed-species trapped-ion modules—each containing one \Sr{88} ion and one \Ca{43} ion—where the \Sr{88} ion serves as the network interface and the \Ca{43} ion provides a robust quantum memory~\cite{drmota_robust_2023}.
We demonstrate the integration of local mixed-species operations with quantum networking by generating bipartite entangled states and multipartite \ac{GHZ} states of up to four qubits.
We also show enhanced storage of remote entanglement using the \Ca{43} memory qubits, preserving entanglement between the modules for more than \SI{10}{\second}, which greatly exceeds the entanglement generation timescale ($\sim\SI{100}{\milli\second}$).

\section*{Quantum networking modules}
\begin{figure}[t]
	\centering
	\includegraphics[width=85mm]{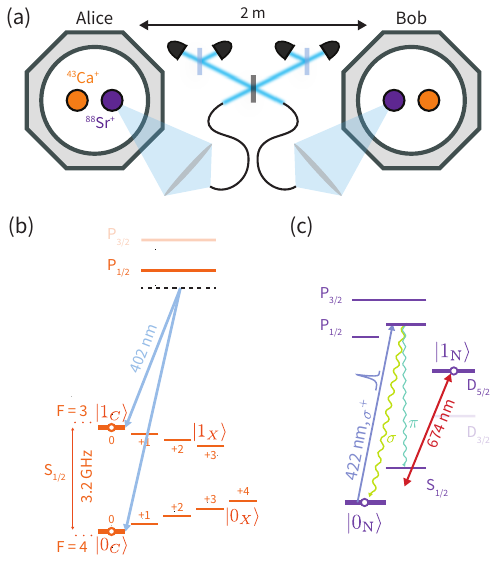}
	\caption{%
		\justifying
		\textbf{Overview of mixed-species trapped-ion quantum networking apparatus.}
		(a) Two modules, Alice and Bob, each co-trap one \Sr{88} and one \Ca{43}. High-numerical-aperture lenses collect single photons from the \Sr{88} ions and couple them into optical fibres. The photons are interfered at a central heralding station, where particular detection patterns herald the generation of entanglement between the \Sr{} ions in each module.
		(b) The ground hyperfine manifold of \Ca{43} provides a robust memory qubit, $\left\{\ket{\down{\cir{}}}, \ket{\up{\cir{}}}\right\}$, and an auxiliary qubit $\left\{\ket{\down{\aux{}}}, \ket{\up{\aux{}}}\right\}$. Transitions between the states within this manifold are driven using a pair of \SI{402}{\nano\meter} Raman beams.
        (c) The \Sr{88} ion provides a \network{} qubit, $\left\{\ket{\down{\net{}}}, \ket{\up{\net{}}}\right\}$, which is connected by the optical \SI{674}{\nano\meter} quadrupole transition. Ion-photon entanglement is generated via the rapid excitation of the \Sr{} ion and subsequent decay via the \SI{422}{\nano\meter} transition, generating entanglement between the polarisation state of the spontaneously emitted photon and the ground Zeeman states of the \Sr{88} ion.
	}
	\label{fig:figure1}
\end{figure}

The quantum networking apparatus comprises two trapped-ion modules connected by a quantum optical link, as depicted in \fig{fig:figure1}(a).
Each module comprises an ultra-high vacuum chamber containing a microfabricated surface-electrode Paul trap, which is used to co-trap one \Sr{88} and one \Ca{43} ion (see Supplementary Information \ref{sup:sec:modules}).
A $\sim$\SI{0.5}{\milli\tesla} magnetic field sets the quantisation axis and lifts the Zeeman state degeneracies, enabling the selection of pairs of energy states to form qubits, as outlined in \fig{fig:figure1}(b) and (c).
Specifically, \Sr{88} provides an optical \emph{network} qubit, $\Q{\net{}}=\{\ket{\down{\net{}}}\equiv\ket{\S{1/2}, m_J=-\tfrac{1}{2}}, \ket{\up{\net{}}}\equiv\ket{\D{5/2}, m_J=-\tfrac{3}{2}}\}$.
The ground hyperfine manifold of \Ca{43} provides a long-lived \emph{circuit} qubit, $\Q{\cir{}}=\{\ket{\down{\cir{}}}\equiv\ket{F=4, m_F=0}, \ket{\up{\cir{}}}\equiv\ket{F=3, m_F=0}\}$, which has a first-order magnetic-field sensitivity of \SI{122}{\kilo\hertz\per\milli\tesla} at \SI{0.5}{\milli\tesla}, roughly two orders of magnitude lower than that of the network qubit.
In addition, an \emph{auxiliary} qubit is defined in the ground hyperfine manifold of \Ca{43}, $\Q{\aux{}}=\{\ket{\down{\aux{}}}\equiv\ket{F=4, m_F=+4}, \ket{\up{\aux{}}}\equiv\ket{F=3, m_F=+3}\}$.
The auxiliary qubit is used in the state preparation and measurement of the circuit qubit, as well as for mediating mixed-species quantum logic.

Within each module, laser beams are used for cooling, state preparation, coherent manipulation, and fluorescence detection of each species of ion.
Single-qubit rotations of the $\Q{\cir{}}$ and $\Q{\aux{}}$ qubits are implemented using a pair of co-propagating \SI{402}{\nano\meter} Raman beams; single-qubit rotations of the optical network qubit, $\Q{\net{}}$, are implemented using a \SI{674}{\nano\meter} laser to drive the quadrupole transition connecting the qubit states, as shown in \fig{fig:figure1}.
To perform mixed-species entangling gates between the two ions, we follow \authorcite{ hughes_benchmarking_2020} in which the interference pattern generated by a pair of non-co-propagating Raman beams gives rise to a $\sigma_z\otimes\sigma_z$-type spin-dependent force~\supplementary{sup:ssec:two_qubit_gates}.
This spin-dependent force is used to implement geometric phase gates between the $\Q{\net{}}$ and $\Q{\aux{}}$ qubits.
As this gate mechanism does not couple to the $\Q{\cir{}}$ qubit, we require the ability to coherently inter-convert between the $\Q{\cir{}}$ and $\Q{\aux{}}$ qubits.
This conversion is performed using the co-propagating Raman beams to coherently address the transitions within the ground hyperfine manifold of \Ca{}~\supplementary{sup:ssec:hyperfine_transfer}.

\begin{figure*}[t]
	\centering
	\includegraphics[width=170mm]{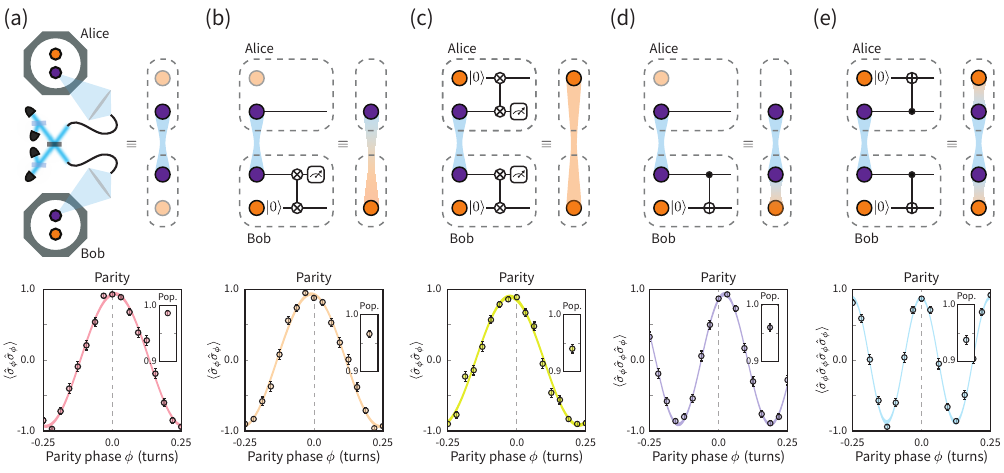}
	\caption{%
		\justifying
		\textbf{Partial state tomography of the mixed-species remotely entangled states.}
		(a) \enquote{Raw} \Sr{}-\Sr{} entanglement.
		(b) \Sr{}-\Ca{} entanglement, created using an error-detected \iswap{}.
		(c) \Ca{}-\Ca{} entanglement, created using error-detected iSWAPs.
		(d) 3-qubit \ac{GHZ} state, created using a \cnot{} gate.
		(e) 4-qubit \ac{GHZ} state, created using \cnot{} gates.
        Parity measurements are shown below each circuit, with insets displaying the population measurements.
        The shaded regions represents the \SI{95}{\percent} confidence interval of the fit given by \eq{eq:parity_fringe}.
        The error bars on the measurements indicate one standard deviation.
        The inferred remote entanglement fidelities are given in \tab{tab:summary}.
	}
	\label{fig:figure2}
\end{figure*}

\begin{table*}[t]
	\centering
	\input{Figures/Summary_Table}
	\caption{%
		\justifying
		\textbf{Summary of mixed-species remotely entangled states characterised in this work.}
        This table includes the observed entanglement fidelities extracted from both partial and full state tomography.
        The density matrices extracted from full state tomography can be found in the supplementary material~\supplementary{supp:sec:tomography}.
    }
	\label{tab:summary}
\end{table*}

Complementing the ability to perform local quantum operations with the two ions in each module, another critical tool is the ability to establish remote entanglement between the two network qubits in separate modules.
This is achieved using a two-photon try-until-success protocol, outlined in \authorcite{stephenson_high-rate_2020}.
Each entanglement generation attempt begins with the preparation of the \Sr{} ion in each module in the ${\ket{\downarrow}=\ket{\S{1/2}, m_J=-\frac{1}{2}}}$ state in the ground Zeeman manifold using approximately \SI{320}{\nano\second} of optical pumping.
A $\sim$\SI{10}{\pico\second} pulse of $\sigma^+$ polarised \SI{422}{\nano\meter} light excites the \Sr{} ion to the $\ket{\P{1/2},m_J=+\frac{1}{2}}$ state, which has a lifetime of \SI{7}{\nano\second}.
The excited state decays to the ground Zeeman states, resulting in the spontaneous emission of a \SI{422}{\nano\meter} photon~\footnote{There is also a \SI{5}{\percent} branching ratio for decays to the $\D{3/2}$ manifold, however these result in an unsuccessful entanglement attempt and so can be neglected.}.
The spontaneously-emitted photons are collected from each module using high-numerical-aperture imaging systems, which couple the photons into single-mode optical fibres.
These fibres bring the spontaneously-emitted photons from each module to a central heralding station, where a projective Bell state measurement is performed on the photons.
Particular measurement outcomes herald the successful generation of entanglement between the Zeeman states of the \Sr{} ions in the two modules.
Finally, a \SI{674}{\nano\meter} pulse maps one of the Zeeman states into the $\D{5/2}$ manifold, thereby establishing the remote entanglement of \network{} qubits.

\section*{Characterisation of entangled states}
In this work, we combine the ability to generate remote entanglement between trapped-ion modules with local mixed-species quantum logic to create higher-dimensional entangled states.
In particular, we consider $N$-particle states of the form
\begin{equation}\label{eq:target_state}
	\ket{\psi} = \frac{\ket{\down{}}^{\otimes N}+e^{i\varphi}\ket{\up{}}^{\otimes N}}{\sqrt{2}},
\end{equation}
where for $N=2$, this is the $\ket{\Phi^+}$ Bell-state, and for $N>2$, this is the $N$-particle \ac{GHZ} state, up to a phase $\varphi$.

We characterise the generated  multipartite states using both partial~\cite{sackett_experimental_2000} and full state tomography~\supplementary{sup:ssec:characterisation}.
Full state tomography enables the complete reconstruction of the density matrix representing the quantum state of the ions from a tomographically complete set of single-qubit measurements.
Partial state tomography assumes that the ideal target states have the form given by \eq{eq:target_state}; thus to characterise the fidelity of the produced state to a particular target state, it is sufficient to perform population measurements, i.e., measurements of all qubits in the computational basis to determine ${P=\mathbb{P}\left(\ket{\down{}}^{\otimes N}\right)+\mathbb{P}\left(\ket{\up{}}^{\otimes N}\right)}$, and parity measurements, i.e., measurements of all qubits in the ${\hat{\sigma}_\phi=\cos(\phi)\hat{\sigma}_x+\sin(\phi)\hat{\sigma}_y}$ basis.
Given quantum states of the form of~\eq{eq:target_state}, we anticipate that the expectation values for the parity measurements will be of the form
\begin{equation}\label{eq:parity_fringe}
	\expect{\bigotimes_{n=1}^N\hat{\sigma}_{\phi,n}}=C\cos(N\phi-\varphi),
\end{equation}
where $C$ and $\varphi$ are the parity contrast and phase, respectively.
By performing these population/parity measurements, we can estimate the fidelity to the target state, up to the parity phase offset, as
\begin{equation}
	\mathcal{F}=\frac{P+C}{2}.
\end{equation}

A summary of all the entanglement fidelities and average generation rates for the remotely entangled states generated in this work is given in \tab{tab:summary}.

\subsection*{Bipartite mixed-species entanglement}
We first characterise remotely entangled bipartite states, i.e., states with $N=2$, starting with the remote entanglement of two \network{} qubits, shown in \fig{fig:figure2}(a).
Full state tomography with \num{200000} tomographic measurements of the \Sr{}-\Sr{} ions yielded a density matrix representing a state with an entanglement fidelity of \SI{96.94(9)}{\percent}.
This is consistent with the results of partial state tomography, shown in \fig{fig:figure2}(a), which yielded a fidelity of \srsrparityfidelity{}.
The average entanglement generation success probability was measured to be \num{1.236(3)e-4}.
We suppress excess heating during entanglement generation by interleaving \SI{500}{\micro\second} of entanglement generation attempts with approximately \SI{320}{\micro\second} of Doppler cooling and \SI{500}{\micro\second} of \ac{EIT} cooling.
Including this cooling, we observe a net entanglement generation rate of \SI{39.31(9)}{\per\second}.

We combine this remote entanglement generation between each module's \network{} qubits with local operations to create mixed-species entangled states.
In particular, we use local mixed-species \iswap{} gates to transfer the quantum states of the remotely entangled \network{} qubits to the \Ca{} \auxiliary{} qubits, thereby enabling the generation of remotely entangled \Sr{}-\Ca{} and \Ca{}-\Ca{} states, as shown in \fig{fig:figure2}(b) and (c), respectively.
To enhance the fidelity of this quantum state transfer process, we perform mid-circuit measurements of the \Sr{} ions after the \iswap{} gates to detect errors in the mixed-species gates~\supplementary{sup:sssec:iswap}.
The modules exchange the outcomes of these measurements in real time via a classical communication link connecting the two control systems, allowing them to abort and restart the experiment if either module detects an error.

As with the \Sr{}-\Sr{} entangled state, we characterise the \Sr{}-\Ca{} and \Ca{}-\Ca{} remotely entangled states using both full and partial state tomography.
Full state tomography with \num{10000} tomographic measurements yielded density matrices representing states with entanglement fidelities of \srcafidelitytonearest{} and \cacafidelitytonearest{} for the \Sr{}-\Ca{} and \Ca{}-\Ca{} states, respectively.
The parity/population measurements for the \Sr{}-\Ca{} and \Ca{}-\Ca{} states are shown in \fig{fig:figure2}(b) and \fig{fig:figure2}(c), respectively.
From these measurements, we extract entanglement fidelities of \srcaparityfidelity{} and \cacaparityfidelity{} for the \Sr{}-\Ca{} and \Ca{}-\Ca{} states, respectively.
For these experiments, we interleave \SI{200}{\micro\second} of entanglement generation attempts with \SI{500}{\micro\second} of Doppler cooling and \SI{500}{\micro\second} of \ac{EIT} cooling using the \Sr{} ions since the local mixed-species operations require the ion crystal to be cooled to close to its motional ground state, resulting in an overall reduction of entanglement generation rate.
This cooling sequence allows us to achieve average phonon numbers of the two-qubit gate motional mode of $\bar{n}\sim0.15$ and $\bar{n}\sim0.1$ in Alice and Bob, respectively.

\subsection*{Multipartite mixed-species entanglement}

We go on to generate multipartite entangled states using the mixed-species trapped-ion modules.
As shown in \fig{fig:figure2}(d) and \fig{fig:figure2}(e), these states are generated by first creating a $\ket{\Phi^+}$ Bell state between the \network{} qubits in each module, followed by local mixed-species \cnot{} gates to entangle the \Ca{} \auxiliary{} qubits with the \network{} qubits, creating the states in \eq{eq:target_state} with $N=3$ and $N=4$, respectively.

Full state tomography with \num{10000} tomographic measurements yielded density matrices representing states with entanglement fidelities of \ghzthreefidelity{} and \ghzfourfidelity{} for the 3-qubit and 4-qubit \ac{GHZ} states, respectively.
The parity/population measurements for the 3-qubit and 4-qubit \ac{GHZ} states are shown in \fig{fig:figure2}(d) and \fig{fig:figure2}(e), respectively.
From these measurements, we extract entanglement fidelities of \ghzthreeparityfidelity{} and \ghzfourparityfidelity{} for the 3-qubit and 4-qubit \ac{GHZ} states, respectively.

\begin{figure}[t]
	\centering
	\includegraphics[width=85mm]{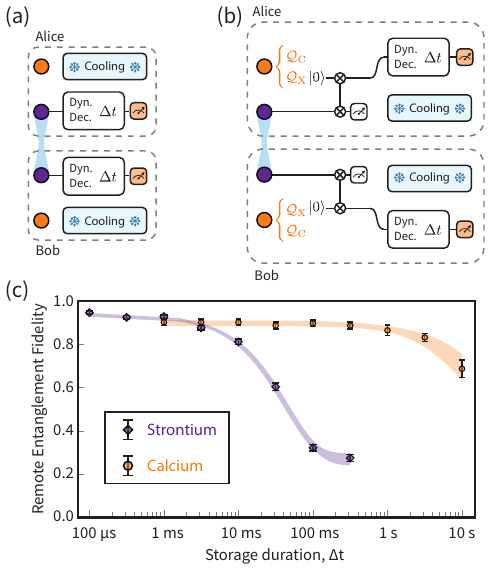}
	\caption{%
		\justifying
		\textbf{Demonstration of the long-lived storage of remote entanglement.}
        Circuits for probing the remote entanglement fidelity as a function of storage duration for (a) two \network{} qubits and (b) two \circuit{} qubits. Sympathetic cooling via the unused ion species and dynamical decoupling is performed to extend the storage capabilities of each qubit.
		(c) Remote entanglement fidelity as a function of storage duration for each qubit. Exponential decay fits (shaded bands) yield decay time constants of \SI{44(3)}{\milli\second} and \SI{14(4)}{\second} for the \network{} and \circuit{} qubits, respectively.
        All error bars indicate one standard deviation.
	}
	\label{fig:figure3}
\end{figure}

\section*{Storage of remote entanglement}
The ability to faithfully preserve quantum information in the modules on timescales longer than the time taken to generate remote entanglement is critical for many quantum networking applications.
In the mixed-species experiments presented so far, the additional cooling required to implement the mixed-species operations means that this timescale is on the order of \SI{100}{\milli\second}.
We have observed coherence times of the \network{} qubit of approximately \SI{8}{\milli\second}; while this can be somewhat extended using dynamical decoupling, it is not sufficient for practical applications.
We therefore utilise the long-lived \circuit{} qubit in \Ca{43} to gain access to a long-lived memory qubit.
In previous work, we demonstrated the integration of this \circuit{} qubit into one of our network modules~\cite{drmota_robust_2023}; we stored the ionic half of an ion-photon Bell pair for up to \SI{10}{\second}, and demonstrated that the storage was robust to network activity, i.e., the simultaneous generation of photons using the co-trapped \Sr{} ion.
Here, we extend this to both modules, thereby enabling the long-lived storage of entanglement between remote matter qubits.

In order to demonstrate the storage enhancement provided by our \circuit{} qubits, we compare the abilities of the \network{} and \circuit{} qubits to store remote entanglement, measuring the fidelity as a function of storage duration.
The experimental circuits used to characterise the storage capabilities of the qubits are shown in \fig{fig:figure3}(a) and (b).
The first step is to generate entanglement between the desired pair of qubits.
For the \network{} qubits, this is achieved as described above.
To generate the remote entanglement between the \circuit{} qubits, we first use error-detected \iswap{}-gates to map the entanglement from the \network{} to the \auxiliary{} qubits~\supplementary{sup:sssec:iswap}, before using a pair of Raman beams to coherently map the quantum information from the \auxiliary{} qubits to the \circuit{} qubits~\supplementary{sup:ssec:hyperfine_transfer}.

With the entanglement created, we store the states for a variable storage duration $\Delta t$.
We deploy \ac{UR} dynamical decoupling~\cite{genov_arbitrarily_2017} to suppress sources of dephasing.
The lengths of the \ac{UR} sequence was adjusted for each storage duration, ranging from 4 pulses (\ac{UR}4) to 48 pulses (\ac{UR}48).
Additionally, over this duration, the ion species not used to store the entanglement is used as a sympathetic coolant to prevent ion loss and mitigate single-qubit rotation errors due to heating.
Finally, we perform parity and population measurements to estimate the fidelity of the entanglement after storage.

The fidelities of the stored entanglement as a function of storage duration for the two types of qubit is shown in \fig{fig:figure3}(c).
For the \network{} qubit, we observe a significant degradation in the stored entanglement for storage durations exceeding $\SI{10}{\milli\second}$.
We attribute this to the relatively high sensitivity of the \network{} qubits to magnetic field fluctuations (\SI{-11}{\mega\hertz\per\milli\tesla}).
Additionally, we observe that the fidelity of the \network{} qubit entanglement decays below \SI{50}{\percent} at storage durations exceeding $\SI{100}{\milli\second}$, indicating the presence of amplitude damping processes such as the decay of the $\ket{\up{\net{}}}$ state due to the finite lifetime of the \Sr{88} metastable $\D{5/2}$ manifold (approximately \SI{390}{\milli\second}~\cite{letchumanan_lifetime_2005}).
We fit an exponential decay to the fidelities, extracting a decay time constant of \SI{44(3)}{\milli\second} for the \network{} qubits.

For the \circuit{} qubits, we observe a marked enhancement of the storage capability, observing a remote entanglement fidelity of \SI{69(4)}{\percent} after \SI{10}{\second}.
Using the same exponential decay model as above, we extract a decay time constant of \SI{14(4)}{\second}.
This significantly exceeds the average time taken to generate entanglement in the mixed-species experiments ($\sim\SI{100}{\milli\second}$).

\section*{Discussion}
In this Letter, we have demonstrated the integration of remote entanglement generation between trapped-ion modules with local mixed-species quantum logic. Specifically, we have realised bipartite entangled states between \Sr{}-\Sr{}, \Sr{}-\Ca{}, and \Ca{}-\Ca{} ion pairs, as well as 3-qubit and 4-qubit multi-species \ac{GHZ} states. All entangled states were generated with fidelities in the range of \SI{91}{\percent} to \SI{97}{\percent}.

We have also shown that remote entanglement can be reliably stored for more than \SI{10}{\second} in the long-lived \circuit{} qubits of \Ca{43}, a duration approximately two orders of magnitude longer than the average entanglement generation time.
This capability is critical for protocols requiring multiple instances of entanglement generation, such as entanglement distillation~\cite{nigmatullin2016minimally}, where storage buffers are essential.

The demonstrated ability to generate, manipulate, and store remotely entangled states across different ion species represents a key enabling technology for future quantum networks. These techniques are complementary to those being developed to operate a quantum network within the \textit{omg} architecture~\cite{allcock_omg_2021, feng_realization_2024, lai_realization_2025}. Heterogeneous entanglement—such as that between different atomic species—can be utilised for advanced applications, including enhanced frequency comparisons in atomic clock networks. In such networks, entangled states engineered across species with complementary sensitivities can yield improved stability and reduced systematic uncertainties~\cite{akerman_atomic_2018}, or enable remote quantum logic spectroscopy~\cite{schmidt2005spectroscopy}.

Moreover, the generation of multipartite entangled states across remote modules opens pathways to enhanced remote sensing protocols~\cite{komar_quantum_2014, van2024utilizing, finkelstein2402universal}, and could serve as a distributed entanglement resource for quantum error correction in networked quantum computing architectures~\cite{gottesman1999quantum, nguyen2021demonstration}.

\input{end_notes.tex}

\input{acronyms}

\bibliography{library}

\clearpage
\onecolumngrid

\section*{Supplementary Information}
\addcontentsline{toc}{section}{Supplementary Material}

\setcounter{section}{0}
\setcounter{figure}{0}
\setcounter{table}{0}
\renewcommand{\thesection}{S\arabic{section}}
\renewcommand{\thesubsection}{S\arabic{section}.\arabic{subsection}}
\renewcommand{\theequation}{S\arabic{equation}}
\renewcommand{\thefigure}{S\arabic{figure}}
\renewcommand{\thetable}{S\arabic{table}}

\twocolumngrid

\input{Supplementary_Material/supplementary_material}

\end{document}

%% file: style.tex
\usepackage{graphicx}
\usepackage{caption}
\usepackage{ragged2e}
\usepackage{svg}
\usepackage{siunitx}
\usepackage{amsmath}
\usepackage{amssymb}
\usepackage[nolist]{acronym}
\usepackage{tabularray}
\UseTblrLibrary{booktabs}
\usepackage{csquotes}
\usepackage{physics}
\usepackage{pifont, xcolor}

\usepackage[hidelinks]{hyperref}

\setcitestyle{sort&compress}



%% file: commands.tex
\newcommand{\Title}{Multipartite Mixed-Species Entanglement over a Quantum Network}
\newcommand{\authorcite}[1]{Ref.~\cite{#1}}

\newcommand{\fig}[1]{Fig.~\ref{#1}}

\newcommand{\tab}[1]{Table~\ref{#1}}
\newcommand{\supplementary}[1]{\hyperref[#1]{(see Supplementary Information)}}
\newcommand{\eq}[1]{Eq.~\ref{#1}}
\newcommand{\sect}[1]{Section~\ref{#1}}
\newcommand{\expect}[1]{\left\langle#1\right\rangle}

\newcommand{\iswap}{iSWAP}
\newcommand{\swap}{SWAP}
\newcommand{\cnot}{CNOT}

\newcommand{\network}{network}
\newcommand{\auxiliary}{auxiliary}
\newcommand{\circuit}{circuit}

\newcommand{\ion}[2]{\mbox{$^{#2}$#1$^+$}}
\newcommand{\Ca}[1]{\ion{Ca}{#1}}
\newcommand{\Sr}[1]{\ion{Sr}{#1}}

\renewcommand{\S}[1]{\text{S}_{#1}}
\renewcommand{\P}[1]{\text{P}_{#1}}
\newcommand{\D}[1]{\text{D}_{#1}}

\newcommand{\down}[1]{0_{#1}}
\newcommand{\up}[1]{1_{#1}}
\newcommand{\Q}[1]{\mathcal{Q}_{#1}}
\newcommand{\net}{\mathrm{N}}
\newcommand{\Qnet}{\Q{\net{}}}
\newcommand{\aux}{\mathrm{X}}
\newcommand{\Qaux}{\Q{\aux{}}}
\newcommand{\cir}{\mathrm{C}}
\newcommand{\Qcir}{\Q{\cir{}}}
\renewcommand{\L}[1]{L\left(#1\right)}
\newcommand{\idn}{\hat{\mathbb{I}}}
\renewcommand{\tr}[1]{\mathrm{tr}\left[#1\right]}
\renewcommand{\r}[2]{\hat{R}_{#1}\left(#2\right)}
\newcommand{\superop}{\hat{\mathcal{S}}}
\newcommand{\kett}[1]{\vert#1\rangle\!\rangle}
\newcommand{\braa}[1]{\langle\!\langle#1\vert}

\newcommand{\dougal}{DM}

\newcommand{\davidN}{DPN}
\newcommand{\ellis}{EMA}

\newcommand{\raghu}{RS}
\newcommand{\gabriel}{GA}
\newcommand{\davidL}{DML}

\newcommand{\rawfidelitytonearest}{\SI{96.94(9)}{\percent}} 
\newcommand{\srcafidelitytonearest}{\SI{94.1(6)}{\percent}}
\newcommand{\cacafidelitytonearest}{\SI{93.1(7)}{\percent}}
\newcommand{\ghzthreefidelity}{\SI{93.1(7)}{\percent}} 
\newcommand{\ghzfourfidelity}{\SI{91.9(8)}{\percent}} 

\newcommand{\srsrparityfidelity}{\SI{96.0(7)}{\percent}}
\newcommand{\srcaparityfidelity}{\SI{95.1(8)}{\percent}}
\newcommand{\cacaparityfidelity}{\SI{92(1)}{\percent}}
\newcommand{\ghzthreeparityfidelity}{\SI{94(1)}{\percent}}
\newcommand{\ghzfourparityfidelity}{\SI{91(1)}{\percent}}

\newcommand{\alicespcirmilli}{\num{4.1(3)}}
\newcommand{\alicemeascirdownmilli}{\num{4.1(4)}} 
\newcommand{\alicemeascirupmilli}{\num{2.7(4)}} 
\newcommand{\alicespauxmilli}{\num{3.1(4)}}
\newcommand{\alicemeasauxdownmilli}{\num{1.77(4)}} 
\newcommand{\alicemeasauxupmilli}{\num{0.357(4)}}
\newcommand{\alicespnetmilli}{\num{4.7(5)}}
\newcommand{\alicemeasnetdownmilli}{$<\num{0.001}$} 
\newcommand{\alicemeasnetupmilli}{\num{1.068(4)}} 

\newcommand{\alicemeasnetavgmilli}{\num{0.534(2)}}
\newcommand{\alicemeasciravgmilli}{\num{3.4(4)}}
\newcommand{\alicemeasauxavgmilli}{\num{1.06(2)}}

\newcommand{\bobspcirmilli}{\num{4.6(3)}}
\newcommand{\bobmeascirdownmilli}{\num{2.7(5)}} 
\newcommand{\bobmeascirupmilli}{\num{1.7(5)}} 
\newcommand{\bobspauxmilli}{\num{3.8(4)}}
\newcommand{\bobmeasauxdownmilli}{\num{1.02(2)}} 
\newcommand{\bobmeasauxupmilli}{$<\num{0.001}$} 
\newcommand{\bobspnetmilli}{\num{5.0(5)}}
\newcommand{\bobmeasnetdownmilli}{$<\num{0.001}$} 
\newcommand{\bobmeasnetupmilli}{\num{1.085(4)}} 

\newcommand{\bobmeasnetavgmilli}{\num{0.543(2)}}
\newcommand{\bobmeasciravgmilli}{\num{2.2(5)}}
\newcommand{\bobmeasauxavgmilli}{\num{0.51(1)}}

\newcommand{\aliceiswapfidelity}{\SI{95.9(2)}{\percent}}
\newcommand{\bobiswapfidelity}{\SI{96.0(2)}{\percent}}

\newcommand{\aliceNXtransfer}{\SI{97.8(3)}{\percent}}
\newcommand{\aliceNXtransferED}{\SI{99.0(2)}{\percent}}
\newcommand{\aliceNXtransferEDP}{\SI{2.8(3)}{\percent}}
\newcommand{\bobNXtransfer}{\SI{97.9(2)}{\percent}}
\newcommand{\bobNXtransferED}{\SI{99.0(2)}{\percent}}
\newcommand{\bobNXtransferEDP}{\SI{2.2(3)}{\percent}}

%% file: title-config.tex
\author{D.~Main}
\altaffiliation{Current address: Oxford Ionics, Oxford, OX5 1GN}
\email{dougal.main@oxionics.com}
\author{P.~Drmota}
\author{E.~M.~Ainley}
\author{A.~Agrawal}
 \author{D.~Webb}
\author{S.~Saner}
\author{O.~B\v{a}z\v{a}van}
\author{B. C. Nichol}
\author{R.~Srinivas}
\author{D.~P.~Nadlinger}
\author{G.~Araneda}\email{gabriel.aranedamachuca@physics.ox.ac.uk}
\author{D.~M.~Lucas}

\affiliation{Department of Physics, University of Oxford, Clarendon Laboratory, Parks Road, Oxford OX1 3PU, United Kingdom}

%% file: Figures/Summary_Table.tex
\SetTblrInner{colsep=1pt,rowsep=3pt}
\begin{tblr}{
  width   = \linewidth,
  colspec = {*{9}{X[c]}},
  row{1}  = {font=\bfseries},
  row{2-Z}= {m},
}
	\hline[1pt]
	\hline[1pt]
	\SetCell[c=2]{c} Ions &              & State                                                                                                           & \SetCell[c=2]{c} Fidelity &                          & \iswap{} \newline Error Prob. & Total \newline Success Prob. & Average \newline Rate             \\
	Alice                 & Bob          &                                                                                                                 & \acs*{PST}                & \acs*{FST}               &                               & $\times10^{-4}$              & \si{\per\second} \\
	\hline[1pt]
	\Sr{}                 & \Sr{}        & $\ket{\down{\net{}}\down{\net{}}}+\ket{\up{\net{}}\up{\net{}}}$                                                 & \srsrparityfidelity{}     & \rawfidelitytonearest{}  & $-$                           & \num{1.236(3)}               & \num{39.31(9)}   \\
	\Sr{}                 & \Ca{}        & $\ket{\down{\net{}}\down{\aux{}}}+\ket{\up{\net{}}\up{\aux{}}}$                                                 & \srcaparityfidelity{}     & \srcafidelitytonearest{} & \SI{4.2(2)}{\percent}         & \num{1.03(1)}                & \num{7.14(7)}    \\
	\Ca{}                 & \Ca{}        & $\ket{\down{\aux{}}\down{\aux{}}}+\ket{\up{\aux{}}\up{\aux{}}}$                                                 & \cacaparityfidelity{}     & \cacafidelitytonearest{} & \SI{8.4(3)}{\percent}         & \num{1.24(1)}                & \num{8.6(1)}     \\
	\Sr{}                 & \Sr{}, \Ca{} & $\ket{\down{\net{}}\down{\net{}}\down{\aux{}}}+\ket{\up{\net{}}\up{\net{}}\up{\aux{}}}$                         & \ghzthreeparityfidelity{} & \ghzthreefidelity{}      & $-$                           & \num{1.20(1)}                & \num{8.26(8)}    \\
	\Sr{}, \Ca{}          & \Sr{}, \Ca{} & $\ket{\down{\net{}}\down{\net{}}\down{\aux{}}\down{\aux{}}}+\ket{\up{\net{}}\up{\net{}}\up{\aux{}}\up{\aux{}}}$ & \ghzfourparityfidelity{}  & \ghzfourfidelity{}       & $-$                           & \num{1.44(2)}                & \num{9.9(1)}     \\
	\hline[1pt]
	\hline[1pt]
\end{tblr}

%% file: end_notes.tex
\section*{Acknowledgements}
We thank Chris Ballance, Joe Goodwin, Laurent Stephenson, P\'eter Juh\'asz, and Jake Blackmore for their contributions to the design and construction of the apparatus, Sandia National Laboratories for supplying the ion traps used in this experiment, and the developers of the control system ARTIQ~\cite{ARTIQ}.
\dougal{} acknowledges support from the U.S.\ Army Research Office (ref.\ W911NF-18-1-0340).
\davidN{} acknowledges support from Merton College, Oxford.
\ellis{} acknowledges support from the U.K.\ EPSRC \enquote{Quantum Communications} Hub EP/T001011/1.
\raghu{} acknowledges funding from an EPSRC Fellowship EP/W028026/1 and Balliol College, Oxford.
\gabriel{} acknowledges support from Wolfson College, Oxford and Cisco.
This work was supported by the U.K.\ EPSRC \enquote{Quantum Computing and Simulation} Hub EP/T001062/1.
\raghu{} is partially employed by Oxford Ionics Ltd.
\davidL{} consults for Quantinuum.
The remaining authors declare no competing interests.

%% file: acronyms.tex
\begin{acronym}
	\acro{GHZ}{Greenberger-Horne-Zeilinger}
	\acro{PST}{partial state tomography}
	\acro{FST}{full state tomography}
	\acro{EIT}{electromagnetically induced transparency}
	\acro{UR}{universally robust}
	\acro{RBM}{randomised benchmarking}
	\acro{OOP}{out-of-phase}
	\acro{POVM}{positive operator-valued measure}
\end{acronym}

%% file: Supplementary_Material/supplementary_material.tex
\section{Modules}\label{sup:sec:modules}
Our quantum networking apparatus comprises two trapped-ion modules, Alice and Bob, which are connected via optical fibre links.
Each module comprises a room-temperature micro-fabricated surface Paul trap fabricated by Sandia National Laboratories, housed in an ultra-high-vacuum chamber.
In Alice, we use the HOA2 trap~\cite{maunz_high_2016}; in Bob, we use the Phoenix trap~\cite{revelle_phoenix_2020} - the successor to the HOA2 trap.
The traps feature two RF electrodes and \num{100} segmented DC control electrodes, enabling precise control over the trapping potentials and the execution of dynamic waveforms for ion shuttling, splitting and merging of ion crystals, and micromotion compensation.
They also include a \SI{60}{\micro\meter} slot in the trap surface, providing optical access from the rear.

Each module has two imaging systems for collecting fluorescence from the ions.
A rear-side imaging system collects \SI{422}{\nano\meter} and \SI{397}{\nano\meter} light through the slot in the trap chip, enabling simultaneous detection of both ion species.
A front-side imaging system, comprising a 0.6 NA lens located outside the vacuum chamber, collects spontaneously-emitted \SI{422}{\nano\meter} photons from the \Sr{} ion at its focal point and couples them into a single-mode optical fibre.

Each module is assigned a \enquote{host} computer, which is used to submit and schedule experiments, communicate with the experimental control system, and collect, analyse, and save experimental data.
Real-time execution of experimental sequences are handled using the ARTIQ control system~\cite{ARTIQ}.
The majority of the hardware comprising this control system is from the Sinara ecosystem.
A digital link connecting the control systems of the two modules provides a classical communication channel, enabling the exchange of information between the modules in real-time.

\section{Local operations}\label{sup:sec:local_ops}

\subsection{Single-qubit gates}\label{sup:ssec:single_qubit_gates}
The ability to coherently manipulate individual qubits is a critical tool for universal quantum computing.
We coherently manipulate the electronic state of the \Sr{} ions using a narrow-linewidth \SI{674}{\nano\meter} laser to address electric quadrupole transitions between sublevels of the $\S{1/2}$ and $\D{5/2}$ manifolds.
In particular, we address the transition between $\ket{\S{1/2}, m_J=-\tfrac{1}{2}}$ and $\ket{\D{5/2}, m_J=-\tfrac{3}{2}}$, which forms the \network{} qubit.
We benchmark the single-qubit rotations of the \network{} qubit using single-qubit \ac{RBM}.
The results for both modules are given in \tab{tab:spam_table}.

\begin{table*}[ht!]
	\centering
	\input{Supplementary_Material/spam_table}
	\caption{%
		\justifying
		\textbf{Summary of quantum operations within each module.}
		Includes state preparation and measurement (SPAM) errors, single-qubit gate errors, and two-qubit \cnot{} gate errors.
	}
	\label{tab:spam_table}
\end{table*}

We coherently manipulate states in the ground hyperfine manifold of the \Ca{} ions using a pair of co-propagating \SI{402}{\nano\meter} Raman beams, detuned from each other by approximately \SI{3.2}{\giga\hertz}.
This choice of geometry allows us to minimise the relative wavevector of the two beams, and therefore the coupling to the ions' motion.
We benchmark the performance of the single-qubit rotations of the \circuit{} and \auxiliary{} qubits using \ac{RBM}; the results are given in \tab{tab:spam_table}.

\subsection{Two-qubit gates}\label{sup:ssec:two_qubit_gates}
The ability to perform logical entangling gates between ions of different species allows us to delegate roles for performing tasks which have diametric requirements, such as storing quantum information and realising a network interface.
We perform these mixed-species entangling gates using light-shift mediated geometric phase gates~\cite{hughes_benchmarking_2020}.

\begin{figure}[t]
	\centering
	\includegraphics[width=85mm]{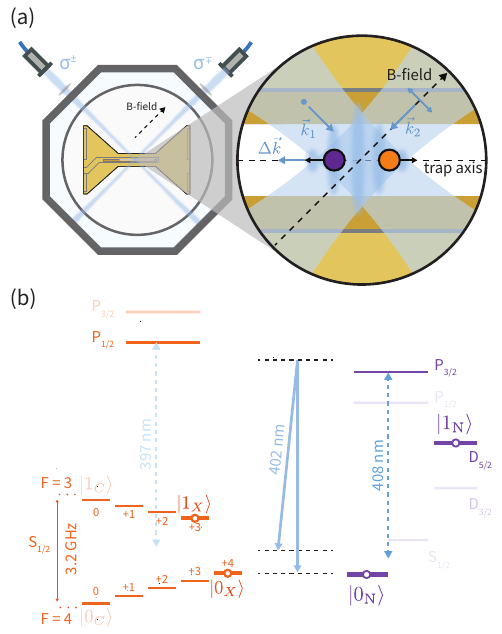}
	\caption{%
		\justifying
		\textbf{Mixed-species entangling gates.}
		(a) Geometry. Two orthogonal Raman beams at $\sim$402 nm, with detuning $\Delta\omega$ with respect each other, drive a light-shift entangling gate between \Ca{43} and \Sr{88}. See details in text.
		(b), Energy level diagram for \Ca{43} and \Sr{88} with the relevant levels used for the two-qubit gate.
	}
	\label{sup-fig:figure2}
\end{figure}

As depicted in \fig{sup-fig:figure2}, the light-shift spin-dependent force is created using a single pair of intersecting \SI{402}{\nano\meter} Raman beams, detuned from one another by $\Delta\omega$, which induce differential light-shifts of the ions' spin states.
At this wavelength, we couple most strongly to the \SI{397}{\nano\metre} $\S{1/2}\leftrightarrow\P{1/2}$ dipole transition in \Ca{} and the \SI{408}{\nano\metre} $\S{1/2}\leftrightarrow\P{3/2}$ dipole transition in \Sr{}, as shown in \fig{sup-fig:figure2}(a).
The $\sim\SI{10}{\tera\hertz}$ detuning from these transitions suppresses unwanted single-photon scattering, while still providing a sufficiently strong driving force with moderate beam intensities.
This configuration is advantageous since we can address both ion species with a single pair of beams.

The interference of the two beams results in a polarisation walking-standing wave propagating along the direction of the wavevector differential $\Delta\vec{k}$, as depicted in \fig{sup-fig:figure2}(b).
The spin states of an ion held in this travelling-standing wave will experience a light-shift oscillating at $\Delta\omega$ with a phase depending on the position of the ion in the travelling-standing wave.
The ions experience this spatially-dependent light-shift as a dipole force which drives motion along $\Delta\vec{k}$, with a sign and magnitude dependent on the spin state of the ion.
By choosing the differential detuning of the Raman beams to be close to the secular frequency of one of the motional modes, i.e., $\Delta\omega=\omega_m+\delta$ where $\abs{\delta}\ll\omega_m$, we off-resonantly drive the motional mode and coherently displace the spin states around approximately circular trajectories in the phase space of the motional mode.

The mass disparity between \Sr{} and \Ca{} results in large asymmetries in the radial mode participations of the ions, making these modes unsuitable for mediating entangling gates.
Instead, we choose the axial \ac{OOP} mode since each ion has a significant participation in the mode, and the \enquote{breathing} character of the ions' motion provides some suppression of the motional heating due to common-mode electric field noise.
The Raman beams are therefore aligned in an orthogonal geometry such that their differential wavevector points in the axial direction and we can maximally drive the axial motion.

The coupling of the spin-dependent force to the spin state depends on the ability to generate differential light shifts between the qubit states.
However, for the \Ca{} \circuit{} qubits, the differential light shift is negligible, and thus the spin-dependent force will drive the states along the same phase-space trajectories, and so cannot be used to generate entanglement.
We therefore drive the entangling gates using the \auxiliary{} qubit instead.
Furthermore, we apply the gate mechanism directly to the \network{} qubit in \Sr{}---rather than the Zeeman ground state qubit, as done in \authorcite{hughes_benchmarking_2020}---which has the advantage of eliminating mapping pulses to convert between the Zeeman and optical qubit at the cost of reduced gate efficiency.

We employ $1^{\mathrm{st}}$-order Walsh modulation~\cite{hayes_reducing_2011, hayes_coherent_2012}, a dynamical error suppression protocol which allows us to coherently suppress certain error processes occuring during the gate and symmetrises the phase space trajectories of the qubit states.
This consists of dividing the gate interaction into two applications of the spin-dependent force, separated by $\hat{\sigma}_x$ rotations of each qubit.
With the Walsh sequence, the asymmetries only result in a further reduction of the gate efficiency that is compensated for by the use of higher laser powers, but the same unitary can be achieved.

\subsubsection{CNOT gate}\label{sup:sssec:cnot}
Through application of the Walsh-modulated spin-dependent force combined with single-qubit unitary operations, we implement a two-qubit \cnot{} gate between the \network{} and \auxiliary{} qubits.
This mixed-species \cnot{} gate is a critical tool, enabling the local entanglement of the \Sr{} and \Ca{} ions for the creation of the \ac{GHZ} states, and also the coherent transfer of quantum information between the two species of ions, discussed in \sect{sup:sssec:iswap}.

We use Raman laser powers of approximately \SI{20}{\milli\watt} per beam, focussed to a beam waist of approximately \SI{15}{\micro\meter}.
With these parameters, we achieve mixed-species \cnot{} gates with a gate duration of \SI{62}{\micro\second} and \SI{58}{\micro\second} in Alice and Bob, respectively.
We characterise the action of the \cnot{} gates using quantum process tomography.
Quantum process tomography provides a complete characterisation of the gate process, enabling the reconstruction of the superoperators representing the actions of the two-qubit gates in each module.
We reconstruct the superoperators for each module from \num{32400} tomographic measurements.
Compared to the ideal \cnot{} gate, we measure average gate fidelities of \SI{97.6(2)}{\percent} and \SI{98.0(2)}{\percent} for Alice and Bob, respectively.

\subsubsection{ISWAP gate}\label{sup:sssec:iswap}
The ability to coherently transfer quantum information from \Sr{} to \Ca{} is an important resource in our trapped-ion modules, enabling the remote entanglement generated across the quantum network to be transferred from the \Sr{} \network{} qubit to the long-lived \Ca{} \circuit{} qubit.

The naive choice for implementing this transfer is the \swap{} gate, which exchanges the quantum states of the participating qubits.
The \swap{} gate requires 3 instances of the \cnot{} gate and therefore significant errors can accumulate, particularly when the transfer is implemented in both modules.
However, in typical scenarios, we only wish to map the state of the \network{} qubit onto one of the \Ca{} qubits, and thus the \Ca{} ion will initially be in a known state, e.g., $\ket{\down{\aux{}}}$.
Hence, it is sufficient to implement the \iswap{} gate, which requires only 2 instances of the \cnot{} gate.

\begin{figure}[t]
	\centering
	\includegraphics[width=85mm]{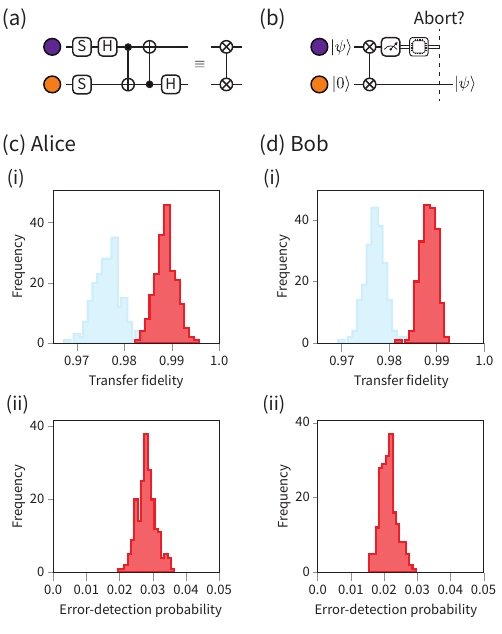}
	\caption{%
		\justifying
		\textbf{\iswap{} gate}
		(a) \iswap{} gate circuit.
		(b) \iswap{} gate circuit with error-detection.
            (c)(i) and (d)(i) show the \iswap{} transfer fidelities without error detection (light blue), and with error detection (red), for Alice and Bob, respectively.
            (c)(ii) and (d)(ii) show the error-detection probabilities, for Alice and Bob, respectively.
	}
	\label{sup-fig:figure3}
\end{figure}

The \iswap{} circuit is shown in \fig{sup-fig:figure3}(a), and implements the unitary
\[
	\hat{U}_{\text{\tiny \iswap{}}}=\begin{pmatrix}
		1 & 0 & 0 & 0 \\
		0 & 0 & i & 0 \\
		0 & i & 0 & 0 \\
		0 & 0 & 0 & 1
	\end{pmatrix}.
\]
Thus, if we let the \Sr{} \network{} qubit be in some arbitrary state $\ket{\psi_{\net{}}}$ and the \Ca{} \auxiliary{} qubit be prepared in the state $\ket{\down{\aux{}}}$, the \iswap{} gate implements the mapping
\begin{equation}\label{eq:iswap_action}
	\ket{\psi_{\net{}}}\ket{\down{\aux{}}}\xrightarrow{\text{\tiny \iswap{}}}\ket{\down{\net{}}}\left(\hat{S}^\dagger\ket{\psi_{\aux{}}}\right)\in\Qnet{}\otimes\Qaux{},
\end{equation}
where $\hat{S}$ is the single-qubit $Z$ rotation through $\pi/4$, given by the unitary matrix $\hat{S}=\mathrm{diag}(1,i)$.
We can then apply an $\hat{S}$-gate to the \auxiliary{} qubit, completing the transfer of the state $\ket{\psi_{\mathrm{N}}}$ from the \Sr{} ion to the \Ca{} ion.

As for the mixed-species \cnot-gate, we characterise the \iswap{} gate using quantum process tomography to reconstruct superoperators, here from a total of \num{32400} measurements.
Compared to the ideal \iswap{}-gate, we measure average gate fidelities of \aliceiswapfidelity{} and \bobiswapfidelity{} for Alice and Bob, respectively, consistent with that expected from the errors in the component gates.

Another important metric is the fidelity with which we transfer quantum information from the \network{} qubit to the \auxiliary{} qubit.
From the reconstructed \iswap{} superoperators, we can predict the action of the \iswap{}-based state transfer, taking into account experimental imperfections.
Such imperfections include imperfect state preparation of the \auxiliary{} qubit and noise in the \iswap{} gate.

Suppose imperfect state preparation of the \auxiliary{} qubit yields the state
\begin{equation}\label{eq:transfer_imperfect_state}
	\hat{\tau}_0=\left(1-\epsilon\right)\ket{\down{}}\bra{\down{}}+\epsilon\ket{\up{}}\bra{\up{}}\in\L{\Qaux{}},
\end{equation}
where $\epsilon$ is the state-preparation error and $\L{\Qaux}$ denotes the space of density matrices for the \auxiliary{} qubit.
Additionally, let the noisy \iswap{} process be represented by the superoperator $\superop{}_{\text{\tiny \iswap{}}}$.
The transfer process, described by
\begin{equation}
	\mathcal{E}_{\net{}\rightarrow\aux{}}:\L{\Qnet{}}\rightarrow\L{\Qaux{}},
\end{equation}
is represented by the superoperator $\superop{}_{\net{}\rightarrow\aux{}}$.
This superoperator is calculated as
\begin{equation}
	\superop{}_{\net{}\rightarrow\aux{}}=\superop{}_S\left(\idn{}_{\Qaux{}}^{\otimes 2}\otimes\braa{\idn{}_{\Q\net{}}}\right)\hat{\mathcal{V}}_2\superop{}_{\text{\tiny \iswap{}}}\hat{\mathcal{V}}_2^\dagger\left(\kett{\hat{\tau}_0}\otimes\idn{}_{\Q\net{}}^{\otimes 2}\right),
\end{equation}
where $\hat{\mathcal{V}}_2$ is the unravelling operation and $\kett{\hat{A}}$ denotes the vectorisation of the operator, $\hat{A}$, as in \authorcite{wood_tensor_2015}.

The ideal transfer process is effectively the identity operation
\begin{equation}
	\mathcal{E}_{\net{}\rightarrow\aux{}}\left(\hat{\rho}\in\L{\Qnet{}}\right)=\hat{\rho}\in\L{\Qaux{}},
\end{equation}
and thus we can calculate the transfer fidelity as in~\cite{wood_tensor_2015},
\begin{equation}\label{eq:transfer_fidelity}
	\mathcal{F}_{\net{}\rightarrow\aux{}}=\frac{d+\tr{{\superop{}_{\net{}\rightarrow\aux{}}}}}{d\left(d+1\right)},
\end{equation}
where $d=\mathrm{dim}\left(\Qaux{}\right)=\mathrm{dim}\left(\Qnet{}\right)=2$.
Histograms showing the spread of transfer fidelities, calculated from resampled \iswap{} gate attempts, are shown for Alice and Bob in \fig{sup-fig:figure3}(c)(i) and (d)(i), respectively.
We infer transfer fidelities of \aliceNXtransfer{} and \bobNXtransfer{} for Alice and Bob, respectively.

Recalling \eq{eq:iswap_action}, since the \auxiliary{} qubit was (nominally) initially prepared in the computational state $\ket{\down{\aux{}}}$, the \network{} qubit will (nominally) be left in the same state, $\ket{\down{\net{}}}$.
As a result of errors in the \iswap{} gate or imperfect state preparation of the \auxiliary{} qubit, the \network{} qubit will occasionally be left in the \enquote{wrong} state.
Therefore, following the \iswap{}-gate, we perform a mid-circuit measurement of the \network{} qubit: if we measure the \network{} qubit in the state $\ket{\down{\net{}}}$, we have projected the system into the $\ket{\down{\net{}}}\bra{\down{\net{}}}\otimes\idn{}_{\aux{}}$ subspace, and thus have effectively projected out part of the \iswap{} gate error.
If the measurement outcome yields the state $\ket{\up{\net{}}}$, we simply restart the protocol.
We perform this measurement and decision branching in real time.

The circuit for this error-detected \iswap{} transfer is shown in \fig{sup-fig:figure3}(b).
As before, we can use the reconstructed \iswap{} superoperator to predict the action of the error-detected transfer in the presence of experimental imperfections, such as imperfect state preparation of the \auxiliary{} qubit, noise in the \iswap{} gate, and imperfect measurement of the \network{} qubit.

As in the case of the non-error-detected transfer, we assume the imperfect state preparation of the \auxiliary{} qubit is described by the state in \eq{eq:transfer_imperfect_state} and that the noisy \iswap{} gate is represented by the superoperator $\superop{}_{\text{\tiny \iswap{}}}$.
After the transfer, we make a measurement of the \network{} qubit, and abort the experiment if we do not observe the qubit to be in the state $\ket{\down{\net{}}}$.
The \ac{POVM} corresponding to a measurement of the \network{} qubit yielding the outcome $\ket{\down{\net{}}}$ is given by
\begin{equation}
	\hat{M}_{\down{}}=\left(1-\epsilon_0\right)\ket{\down{\net{}}}\bra{\down{\net{}}}+\epsilon_1\ket{\up{\net{}}}\bra{\up{\net{}}}.
\end{equation}
We calculate the superoperator representing the error-detected transfer process as
\begin{equation}
	\superop{}'_{\net{}\rightarrow\aux{}}=\superop{}_S\left(\idn{}_{\Qaux{}}^{\otimes 2}\otimes\braa{\hat{M}_{\down{}}}\right)\hat{\mathcal{V}}_2\superop{}_{\text{\tiny \iswap{}}}\hat{\mathcal{V}}_2^\dagger\left(\kett{\hat{\tau}_0}\otimes\idn{}_{\Q\net{}}^{\otimes 2}\right).
\end{equation}
Note that since we have introduced an abort condition, the calculated process is no longer trace-preserving.
Suppose the \network{} qubit is prepared in some state $\hat{\rho}$ that we would like to transfer to the \auxiliary{} qubit, then the probability that an error is detected, denoted $p$, can be calculated as
\begin{equation}
	p=1-\tr{\mathcal{E}'_{\net{}\rightarrow\aux{}}\left(\hat{\rho}\right)}.
\end{equation}
Note that, in general, this probability depends non-linearly on the input state, $\hat{\rho}$, and thus it is not possible to normalise the process.

We calculate the average gate fidelity for the error-detected transfer using a Monte-Carlo approach.
We sample $N$ input states $\ket{\psi_i}$ from the Haar measure and compute the average transfer fidelity as
\begin{equation}\label{eq:ed_transfer_fidelity}
	\bar{\mathcal{F}}_{\net{}\rightarrow\aux}=\frac{1}{N}\sum_{i=1}^N\frac{\bra{\psi_i}\mathcal{E}'_{\net{}\rightarrow\aux{}}\left(\ket{\psi_i}\bra{\psi_i}\right)\ket{\psi_i}}{\tr{\mathcal{E}'_{\net{}\rightarrow\aux{}}\left(\ket{\psi_i}\bra{\psi_i}\right)}}.
\end{equation}
We calculate the average error-detection probability, $\bar{p}$, in similar fashion, as
\begin{equation}\label{eq:ed_probability}
	\bar{p}=1-\frac{1}{N}\sum_{i=1}^N\tr{\mathcal{E}'_{\net{}\rightarrow\aux{}}\left(\ket{\psi_i}\bra{\psi_i}\right)}.
\end{equation}
Histograms showing the spread of average transfer fidelities, $\bar{\mathcal{F}}_{\net{}\rightarrow\aux}$, and error-detection probabilities, $\bar{p}$, as calculated from \iswap{} gates reconstructed from resampled datasets, are shown in \fig{sup-fig:figure3}(c) and (d) for Alice and Bob, respectively.
The results for Alice (Bob) indicate an average transfer fidelity of $\bar{\mathcal{F}}_{\net{}\rightarrow\aux}=\aliceNXtransferED{}$ ($\bar{\mathcal{F}}_{\net{}\rightarrow\aux}=\bobNXtransferED{}{}$) and an average error-detection probability of $\bar{p}=\aliceNXtransferEDP{}$ ($\bar{p}=\bobNXtransferEDP{}$).

\subsection{Hyperfine transfer}\label{sup:ssec:hyperfine_transfer}
Since the \circuit{} qubit does not participate in the mixed-species gate, the gate interaction is performed between the \network{} and \auxiliary{} qubits.
Consequently, we require the ability to coherently inter-convert between the \circuit{} and \auxiliary{} qubit before and after the local operations.
This mapping is performed using the Raman beams to coherently address the transitions within the ground hyperfine manifold of \Ca{43}.

As in \authorcite{main_distributed_2025}, the transfer of the \circuit{} qubit to the \auxiliary{} qubit begins with the mapping of the state $\ket{\down{\cir{}}}$ to the state $\ket{\down{\aux{}}}$.
However, due to the near degeneracy of the transition $\mathcal{T}_0: \ket{\down{\cir{}}}\leftrightarrow\ket{F=3, M_F=+1}$ and the transition $\mathcal{T}_1:\ket{\up{\cir{}}}\leftrightarrow\ket{F=4, M_F=+1}$, separated by only $\Delta f\approx\SI{15}{\kilo\hertz}$, it is not possible to map the $\ket{\down{\cir{}}}$ state out of the \circuit{} qubit without off-resonantly driving population out of the $\ket{\up{\cir{}}}$ state.
We suppress this off-resonant excitation using a composite pulse sequence comprising three pulses resonant with the $\mathcal{T}_0$ transition, with pulse durations equal to the $2\pi$-time of the $\mathcal{T}_1$ transition, and phases optimised to minimise the off-resonant excitation.
This pulse sequence allows us to simultaneously perform a $\pi$-pulse on the $\mathcal{T}_0$ transition and an identity on the off-resonantly-driven $\mathcal{T}_1$ transition.
Raman $\pi$-pulses are then used to complete the mapping to the $\ket{\down{\aux{}}}$ state.
Another sequence of Raman $\pi$-pulses coherently maps $\ket{\up{\cir{}}}\rightarrow\ket{\up{\aux{}}}$, thereby completing the transfer of the \circuit{} qubit to the \auxiliary{} qubit, $\Qcir{}\rightarrow\Qaux{}$.
To implement the mapping $\Qaux{}\rightarrow\Qcir{}$ the same pulse sequence is applied in reverse.

We characterise our $\Qcir\leftrightarrow\Qaux$ transfer sequence by performing a modification of single-qubit \ac{RBM}, in which we alternate Clifford operations on the $\Qcir$ and $\Qaux$ qubits.
We measure an error per transfer of \num{3.8(2)e-3} (\num{2.6(1)e-3}) for Alice (Bob).

\section{Quantum State Tomography}\label{sup:ssec:characterisation}
To characterise the states generated in the experiments described in the main text, we use two methods: full state tomography and partial state tomography.

\subsection{Full state tomography}
A density matrix, $\hat{\rho}\in\L{\Q{}}$, provides a complete description of the quantum state of a system.
Quantum state tomography is the process by which measurements of a number of copies of a state are used to reconstruct the density matrix describing the state of a system.

Suppose we have $N$ of copies of some unknown quantum state that we can measure using some measurement apparatus.
Our measurement apparatus has a number of different measurement settings, labelled by the index $i$, and each measurement setting has a set of possible outcomes, labelled by the index $j$.
The measurement outcomes for a particular measurement setting are associated with the \acp{POVM}, $\hat{M}_{ij}$, satisfying $\sum_j\hat{M}_{ij}=\hat{\idn{}}$, such that the probability that a measurement of the state using the measurement setting $i$ yields the outcome $j$ is given by
\begin{equation}\label{eq:likelihood}
	p_{ij}=\tr{\hat{M}_{ij}\hat{\rho}}.
\end{equation}
Suppose we make measurements of $N$ copies of our state using different measurement settings and count the number of occurrences, $n_{ij}$, of each outcome for each measurement setting.
The likelihood of obtaining a particular dataset, $\{n_{ij}\}$, given the state $\hat{\rho}$, can be written as
\begin{equation}
	\mathcal{L}\left(\hat{\rho}\right)=\Pi_{ij}\left(\tr{\hat{M}_{ij}\hat{\rho}}\right)^{n_{ij}}.
\end{equation}
Therefore, we can find an estimate for the density matrix describing the state of the system by finding the density matrix, $\hat{\rho}$, which maximises the likelihood in \eq{eq:likelihood}.

After performing the tomographic measurements and collecting the raw data, we find the density matrix that maximises the likelihood by following an iterative diluted maximum-likelihood estimation algorithm developed in \authorcite{rehacek_diluted_2007}.

For a system comprising $N$ qubits, the tomographic measurements are constructed by randomly selecting a measurement setting, $i$, for each qubit.
We perform a measurement with this setting by applying a single-qubit rotation
\begin{align}\label{eq:tomographic rotations}
	\hat{U}_i \in \left\{
	\idn{},~
	\idn{},~
	\r{X}{\frac{\pi}{2}},~
	\r{Y}{\frac{\pi}{2}},~
	\r{X}{-\frac{\pi}{2}},~
	\r{Y}{-\frac{\pi}{2}}
	\right\},
\end{align}
to the $n^\mathrm{th}$ qubit and measuring in the computational basis, $\hat{\sigma}_z$.
This corresponds to a measurement of the observable $\hat{U}_i^\dagger\hat{\sigma}_z\hat{U}_i$.
Note that the set of rotations in \eq{eq:tomographic rotations} corresponds to a tomographically over-complete set of measurements, as we measure along the $\pm X$ and $\pm Y$ directions, which improves robustness to systematic errors.
Note that when selecting a measurement setting, we include the $\idn{}$ element twice, ensuring that we acquire equal statistics along each axis.

Assuming perfect single-qubit rotations and qubit readout, the \acp{POVM} for the $m^\mathrm{th}$ qubit can be written as
\begin{align}
	\hat{M}_{i_m,\down{}} & = \hat{U}_i^\dagger\ket{\down{}}\bra{\down{}}\hat{U}_i, \\
	\hat{M}_{i_m,\up{}}   & = \hat{U}_i^\dagger\ket{\up{}}\bra{\up{}}\hat{U}_i.
\end{align}
Thus, the $N$-qubit \acp{POVM} may be constructed as
\begin{equation}
	\hat{M}_{ij}=\bigotimes_{m=1}^N\hat{M}_{i_m,j_m},
\end{equation}
where the overall measurement setting, $i$, is the collection of single-qubit measurement settings, $\left(i_1, i_2, \dots, i_N\right)$, and the overall measurement outcome, $j$, is the collection of the individual single-qubit measurement outcomes, $\left(j_1, j_2, \dots, j_N\right)$.

Thus far, we have assumed perfect tomographic measurements.
Now, to obtain a more accurate characterisation of the quantum system, we incorporate imperfect qubit readout into our algorithm.
Let $\epsilon_0$ and $\epsilon_1$ denote the readout errors of qubit states, such that $\epsilon_0$ ($\epsilon_1$) corresponds to the probability that a measurement of a qubit prepared in the state $\ket{\down{}}$ ($\ket{\up{}}$) will yield the measurement outcome $\up{}$ ($\down{}$).
We therefore construct the single-qubit tomographic measurement \acp{POVM} as
\begin{align}
	\hat{M}_{i, \down{}} & = (1-\epsilon_{\down{}})\hat{U}_i^\dagger\ket{\down{}}\bra{\down{}}\hat{U}_i+\epsilon_{\up{}}\hat{U}_i^\dagger\ket{\up{}}\bra{\up{}}\hat{U}_i,\\
	\hat{M}_{i, \up{}}   & = (1-\epsilon_{\up{}})\hat{U}_i^\dagger\ket{\up{}}\bra{\up{}}\hat{U}_i+\epsilon_{\down{}}\hat{U}_i^\dagger\ket{\down{}}\bra{\down{}}\hat{U}_i.
\end{align}
Experimentally determined values for the readout errors, $\epsilon_0$ and $\epsilon_1$, for the different qubits are given in \tab{tab:spam_table}.
We neglect errors in the single-qubit rotations as we do not yet have a complete understanding of the structure of these errors, however, we expect this to be $<\num{1e-3}$.

Once we have reconstructed a state $\hat{\rho}$, we can calculate the \emph{fidelity} of the state to some target pure state, $\ket{\psi}$, as
\begin{equation}
	\mathcal{F}\left(\hat{\rho}, \ket{\psi}\bra{\psi}\right)=\bra{\psi}\hat{\rho}\ket{\psi}.
\end{equation}
For the characterisation of bipartite states, we make use of the \emph{entanglement fidelity}, $\mathcal{F}_E$, which is defined as the fidelity to the closest maximally entangled state under local operations,
\begin{equation}
	\mathcal{F}_E=\max_{\hat{U}}\mathcal{F}\left(\hat{U}\hat{\rho}\hat{U}^\dagger, \ket{\Phi^+}\right),
\end{equation}
where $\hat{U}=\hat{U}_1\otimes\hat{U}_2$ and $\ket{\Phi^+}$ is a Bell state.
For the \ac{GHZ} states, we define the entanglement fidelity as the fidelity of the state to the \ac{GHZ} state under local $Z$ rotations, such that for $N$ qubits
\begin{equation}
	\mathcal{F}_E=\max_{\hat{U}}\mathcal{F}\left(\hat{U}\hat{\rho}\hat{U}^\dagger, \ket{\mathrm{GHZ}_N}\right),
\end{equation}
where $\hat{U}=\bigotimes_{i=1}^{N}\r{Z}{\phi_i}$ and $\ket{\mathrm{GHZ}_N}$ is the $N$-qubit \ac{GHZ} state.

\subsection{Partial state tomography}
Full quantum state tomography provides a complete characterisation of an unknown quantum state; however, it is expensive in the number of measurements required to achieve a particular precision.
However, given prior knowledge about what form the state should take, partial state tomography via population and parity measurements can be an inexpensive way to extract information about particular properties of the state.

Let us write the states that we create as the density matrix
\begin{equation}
	\hat{\rho}=\sum_{i, j}\rho_{i,j}\ket{i}\bra{j}\in\L{\Q{}^{\otimes N}},
\end{equation}
where $\ket{i}\in\Q{}^{\otimes N}$.
The fidelity of the created state to the desired state, \eq{eq:target_state}, is given by
\begin{align*}
	\mathcal{F} & =\bra{\psi}\hat{\rho}\ket{\psi}\nonumber                                                                                      \\
	            & =\frac{1}{2}\left[\rho_{0\dots0,0\dots0}+\rho_{1\dots1,1\dots1}+2\abs{\rho_{0\dots0,1\dots1}}\cos\left(\varphi\right)\right],
\end{align*}
where $\varphi=\mathrm{arg}\left(\rho_{0\dots0,1\dots1}\right)$.
Let us now define the population observable, $\hat{P}=\hat{\Pi}_0^{\otimes N}+\hat{\Pi}_1^{\otimes N}$, such that
\begin{equation}
	P=\expect{\hat{P}}=\rho_{0\dots0,0\dots0}+\rho_{1\dots1,1\dots1}.
\end{equation}
Furthermore, define the parity observable, $\hat{\sigma}_{\phi}^{\otimes N}$, where $\hat{\sigma}_{\phi}=\cos\left(\phi\right)\hat{\sigma}_x+\sin\left(\phi\right)\hat{\sigma}_y$, such that for states of the form \eq{eq:target_state}, we have
\begin{equation}
	\expect{\hat{\sigma}_{\phi}^{\otimes N}}=C\cos\left(N\phi-\varphi\right),
\end{equation}
where $C=2\abs{\rho_{0\dots0,1\dots1}}$ and $\varphi$ is the parity phase of the created state.
Thus, if we define the fidelity of the created state to the state in \eq{eq:target_state} up to an arbitrary $Z$-rotation, we find that the fidelity is given by
\begin{equation}
	\mathcal{F}=\frac{P+C}{2}.
\end{equation}

Experimentally, the population measurement can be performed by simply measuring the created state in the computational basis, and computing the expectation value $\expect{\hat{P}}$, i.e., the probability that the qubits were all measured to be in the same state, e.g., $\mathrm{P}_{00}+\mathrm{P}_{11}$ for the 2-qubit case.
The parity measurements can be implemented by first performing $\r{\phi+\pi/2}{\pi/2}$ rotations on each qubit, before measuring in the computational basis.
This corresponds to a measurement of the qubits along the axis at an angle $\phi$ to the $X$-axis in the $XY$-plane of the Bloch sphere, associated with the operator $\hat{\sigma}_{\phi}$.
We thus use these measurement outcomes to calculate the expectation values, $\expect{\hat{\sigma}_\phi^{\otimes N}}$.

\section{Tomographically Reconstructed Density Matrices}\label{supp:sec:tomography}
\begin{figure}[t]
	\centering
	\includegraphics[width=85mm]{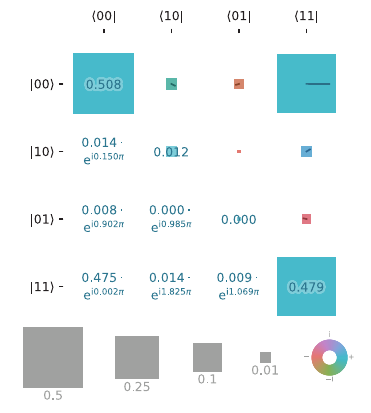}
	\caption{%
		\justifying
		\textbf{Tomographic reconstruction of \Sr{}-\Sr{} remotely entangled state.} The reconstructed density matrix yields an entanglement fidelity of \SI{96.94(9)}{\percent}.
	}
	\label{sup-fig:figure5}
\end{figure}

\begin{figure}[t]
	\centering
	\includegraphics[width=85mm]{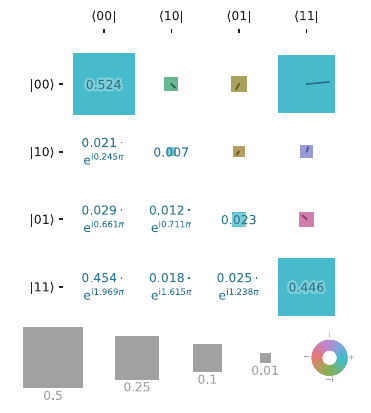}
	\caption{%
		\justifying
		\textbf{Tomographic reconstruction of \Sr{}-\Ca{} remotely entangled state.} The reconstructed density matrix yields an entanglement fidelity of \SI{94.1(6)}{\percent}.
	}
	\label{sup-fig:figure6}
\end{figure}

\begin{figure}[t]
	\centering
	\includegraphics[width=85mm]{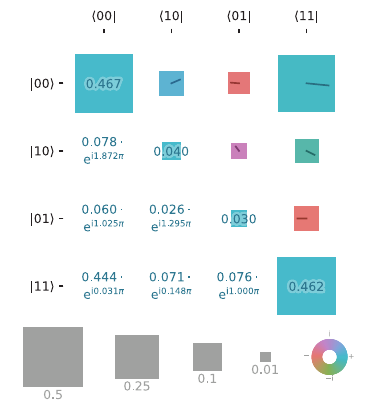}
	\caption{%
		\justifying
		\textbf{Tomographic reconstruction of \Ca{}-\Ca{} remotely entangled state.} The reconstructed density matrix yields an entanglement fidelity of \SI{93.1(7)}{\percent}.
	}
	\label{sup-fig:figure7}
\end{figure}

\begin{figure*}[t]
	\centering
	\includegraphics[width=122.5mm]{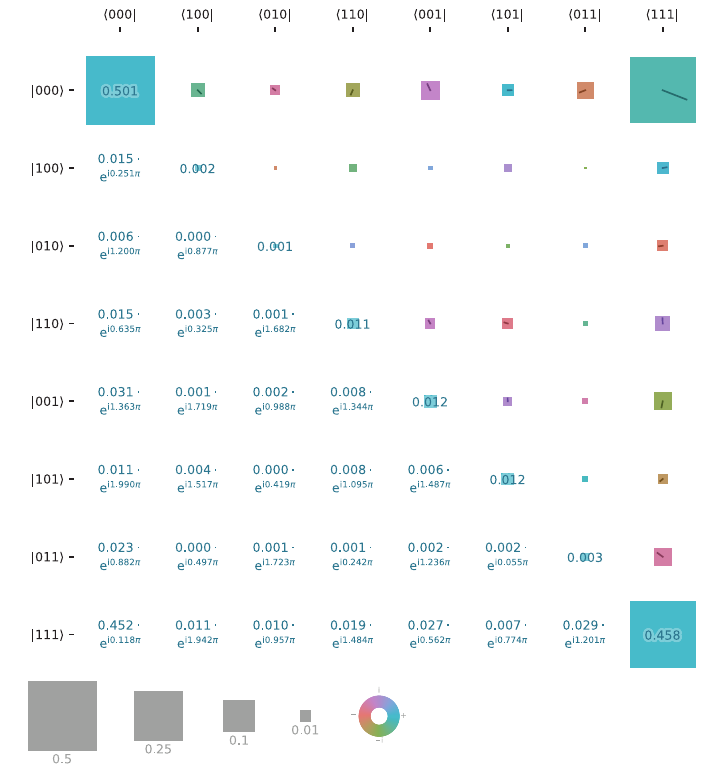}
	\caption{%
		\justifying
		\textbf{Tomographic reconstruction of \Sr{}-\Sr{}-\Ca{} remotely entangled state.} The reconstructed density matrix yields an entanglement fidelity of \SI{93.1(7)}{\percent}.
	}
	\label{sup-fig:figure8}
\end{figure*}

\begin{figure*}[t]
	\centering
	\includegraphics[width=170mm]{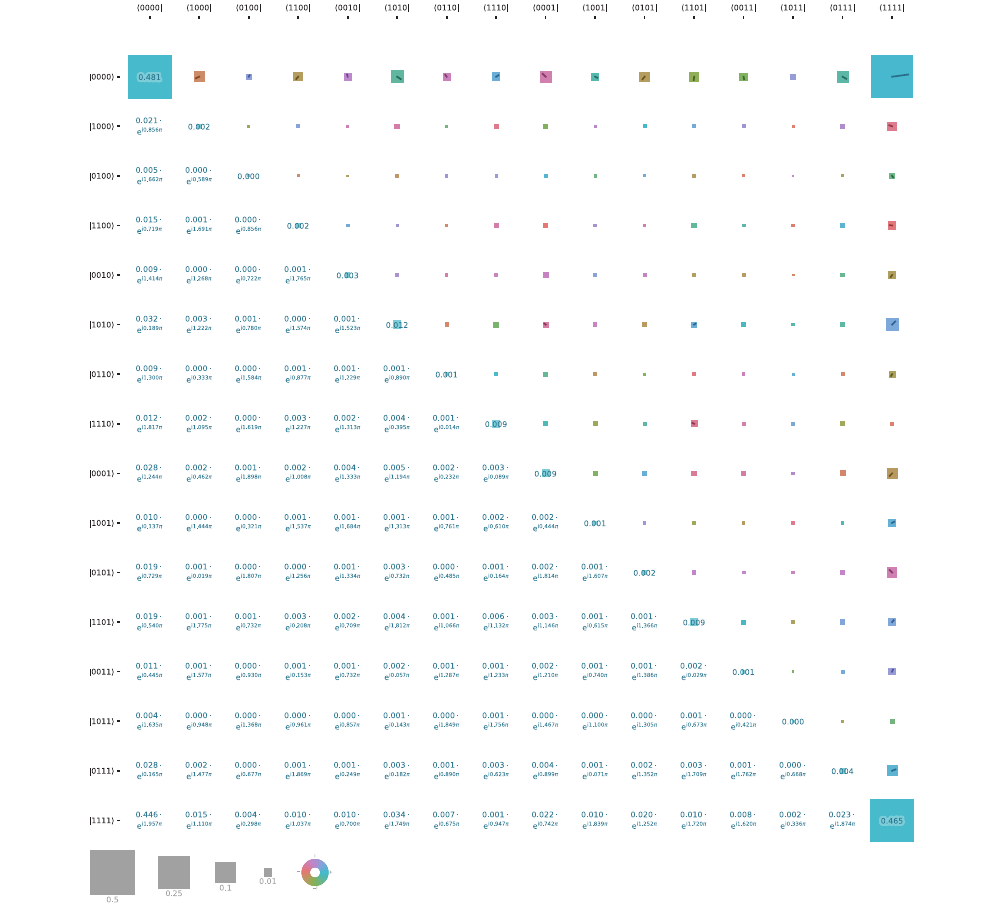}
	\caption{%
		\justifying
		\textbf{Tomographic reconstruction of \Sr{}-\Sr{}-\Ca{}-\Ca{} remotely entangled state.} The reconstructed density matrix yields an entanglement fidelity of \SI{91.9(8)}{\percent}.
	}
	\label{sup-fig:figure9}
\end{figure*}

In Figs.~\ref{sup-fig:figure5}-\ref{sup-fig:figure9}, we provide the reconstructed density matrices for the states generated by the experiments described in the manuscript.
These density matrices yield the entanglement fidelities quoted in the manuscript.

\newpage

%% file: Supplementary_Material/spam_table.tex
\SetTblrInner{colsep=1pt,rowsep=3pt}
\begin{tblr}{
  width   = \linewidth,
  colspec = {*{8}{X[c]}},
  row{1}  = {font=\bfseries},
  row{2-Z}= {m},
}
	\hline[1pt]
	\hline[1pt]
	Module                 & Qubit     & State-prep. ($\times10^{-3}$) & \SetCell[c=3]{c} Readout ($\times10^{-3}$) &                        &                                   & 1Q gate error \newline per Clifford ($\times10^{-4}$) & \cnot{} gate error \newline per gate \\
	                       &           & $\epsilon$                    & $\epsilon_0$                               & $\epsilon_1$           & $\frac{\epsilon_0+\epsilon_1}{2}$ &                                                    &    \\
	\hline[1pt]
	\SetCell[r=3]{c} Alice & $\Qnet{}$ & \alicespnetmilli{}            & \alicemeasnetdownmilli{}                   & \alicemeasnetupmilli{} & \alicemeasnetavgmilli{}           & \num{4.8(3)}                                          & \SetCell[r=3]{c} 0.024(2) \\
	                       & $\Qcir{}$ & \alicespcirmilli{}            & \alicemeascirdownmilli{}                   & \alicemeascirupmilli{} & \alicemeasciravgmilli{}           & \num{1.0(3)}                                       &    \\
	                       & $\Qaux{}$ & \alicespauxmilli{}            & \alicemeasauxdownmilli{}                   & \alicemeasauxupmilli{} & \alicemeasauxavgmilli{}           & \num{1.4(3)}                                       &    \\
	\hline[1pt]
	\SetCell[r=3]{c} Bob   & $\Qnet{}$ & \bobspnetmilli{}              & \bobmeasnetdownmilli{}                     & \bobmeasnetupmilli{}   & \bobmeasnetavgmilli{}             & \num{9.8(3)}                                          &  \SetCell[r=3]{c} 0.020(2)\\
	                       & $\Qcir{}$ & \bobspcirmilli{}              & \bobmeascirdownmilli{}                     & \bobmeascirupmilli{}   & \bobmeasciravgmilli{}             & \num{1.2(4)}                                       &    \\
	                       & $\Qaux{}$ & \bobspauxmilli{}              & \bobmeasauxdownmilli{}                     & \bobmeasauxupmilli{}   & \bobmeasauxavgmilli{}             & \num{1.2(3)}                                       &    \\
	\hline[1pt]
	\hline[1pt]
\end{tblr}